\documentclass{moriond}
\pdfoutput=1
\usepackage{subfigure}

\def\be{\begin{equation}}
\def\ee{\end{equation}}
\def\bea{\begin{eqnarray}}
\def\eea{\end{eqnarray}}

\def\ttbar{\ensuremath{t\bar{t}}}
\def\pt{\ensuremath{p_{T}}}

\newcommand{\Vlr}{V_{\mathrm{L,R}}}
\newcommand{\Glr}{g_{\mathrm{L,R}}} 
\newcommand{\ProjR}{P_{\mathrm{R}}}
\newcommand{\ProjL}{P_{\mathrm{L}}}
\newcommand{\ProjLR}{P_{\mathrm{{L,R}}}}
\def\vl{\ensuremath{V_{\mathrm{L}}}}
\def\vr{\ensuremath{V_{\mathrm{R}}}}
\def\gr{\ensuremath{g_{\mathrm{R}}}}
\def\gl{\ensuremath{g_{\mathrm{L}}}}
\def\FL{\ensuremath{F_{\mathrm{L}}}}
\def\FR{\ensuremath{F_{\mathrm{R}}}}
\def\F0{\ensuremath{F_{0}}}
\def\mW{\ensuremath{m_{W}}}
\def\mlb{\ensuremath{m_{\ell b}}}
\def\mt{\ensuremath{m_{\mathrm{ top}}}}

\newcommand{\Photo}{\includegraphics[height=35mm]{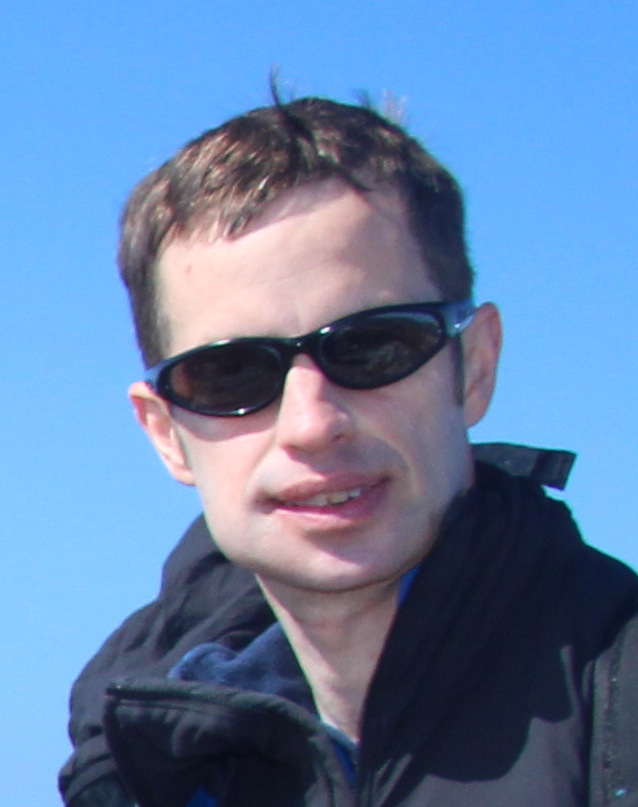}}

\begin{document}
\vspace*{4cm}
\title{Top quark properties measurements at the LHC}

\author{Mark Owen\\
On behalf of the ATLAS and CMS Collaborations}

\address{School of Physics and Astronomy, University of Glasgow,\\
Glasgow, G12 8QQ, UK}

\maketitle\abstracts{
Highlights of measurements of the properties of the top quark at the LHC are presented.
The measurements probe a range of the properties of the top quark,
including the structure of the $Wtb$~vertex, the top-$Z$~coupling and the top-quark
mass.
The results are compared to Standard Model predictions and in some
cases limits on physics beyond the Standard Model are also extracted
in the context of effective field theory models.
The measurements use data collected by the ATLAS and CMS experiments
during $pp$~collisions at a centre-of-mass energy of $8$~or $13$~TeV.
}

\section{Introduction}

The top quark is the heaviest fundamental particle discovered to date
and it decays before it has a chance to hadronise. These characteristics
not only allow for precision tests of the Standard Model (SM), but
also open a potential window to physics beyond the SM.
A selection of recent measurements from the ATLAS~\cite{ATLASdet} and CMS~\cite{CMSdet} collaborations are discussed below\footnote{For
a full list of top properties measurements, please see the public websites of the collaborations:
\url{http://cms-results.web.cern.ch/cms-results/public-results/publications/TOP/index.html},
\url{http://cms-results.web.cern.ch/cms-results/public-results/preliminary-results/TOP/index.html}
and \url{https://twiki.cern.ch/twiki/bin/view/AtlasPublic/TopPublicResults}.
}.
The measurements use both top-quark pair and single-top quark production modes.
For the analyses using $\ttbar$~production, two decay modes with low background rates are utilised:
the lepton-plus-jets decay mode,
where one $W$~boson decays leptonically and the other decays into a pair of quarks
and the dilepton decay mode, where both $W$~bosons decays leptonically.

\section{The $Wtb$~vertex}

In the SM the top quark is predicted to decay almost exclusively into a $W$~boson and $b$-quark.
The decay time of the top quark is shorter than the characteristic time for hadronisation and this
means the top quark provides a unique window to observe the properties of a bare quark.
The decay products of the top quark can therefore be used to probe the nature of the $Wtb$~vertex.

\subsection{W boson polarisation}

The $W$~bosons produced in top decays can be either left-handed, right-handed or
longitudinally polarised. The corresponding fractions ($\FL$, $\FR$~and $\F0$) are well predicted
in the SM, however the presence of new physics in the $Wtb$~vertex could result in fractions
different to the SM predictions. Experimentally, these fractions can be accessed by measuring the
helicity angle $\theta^*$~between the charged lepton or down-type quark and the direction
of the top quark in the rest frame of the $W$~boson. The distribution for the cosine of the helicity
angle depends on the polarisation fractions according to:
\begin{equation}
\frac{1}{\Gamma}\frac{d \Gamma}{d \cos \theta^*} = \frac{3}{8}(1 - \cos \theta^*)^2 \FL + \frac{3}{4}(\sin^2 \theta^*)\F0 + \frac{3}{8}(1 + \cos \theta^*)^2 \FR.
\end{equation}

\begin{figure}
\centering
\subfigure[]{
\includegraphics[width=0.4\linewidth]{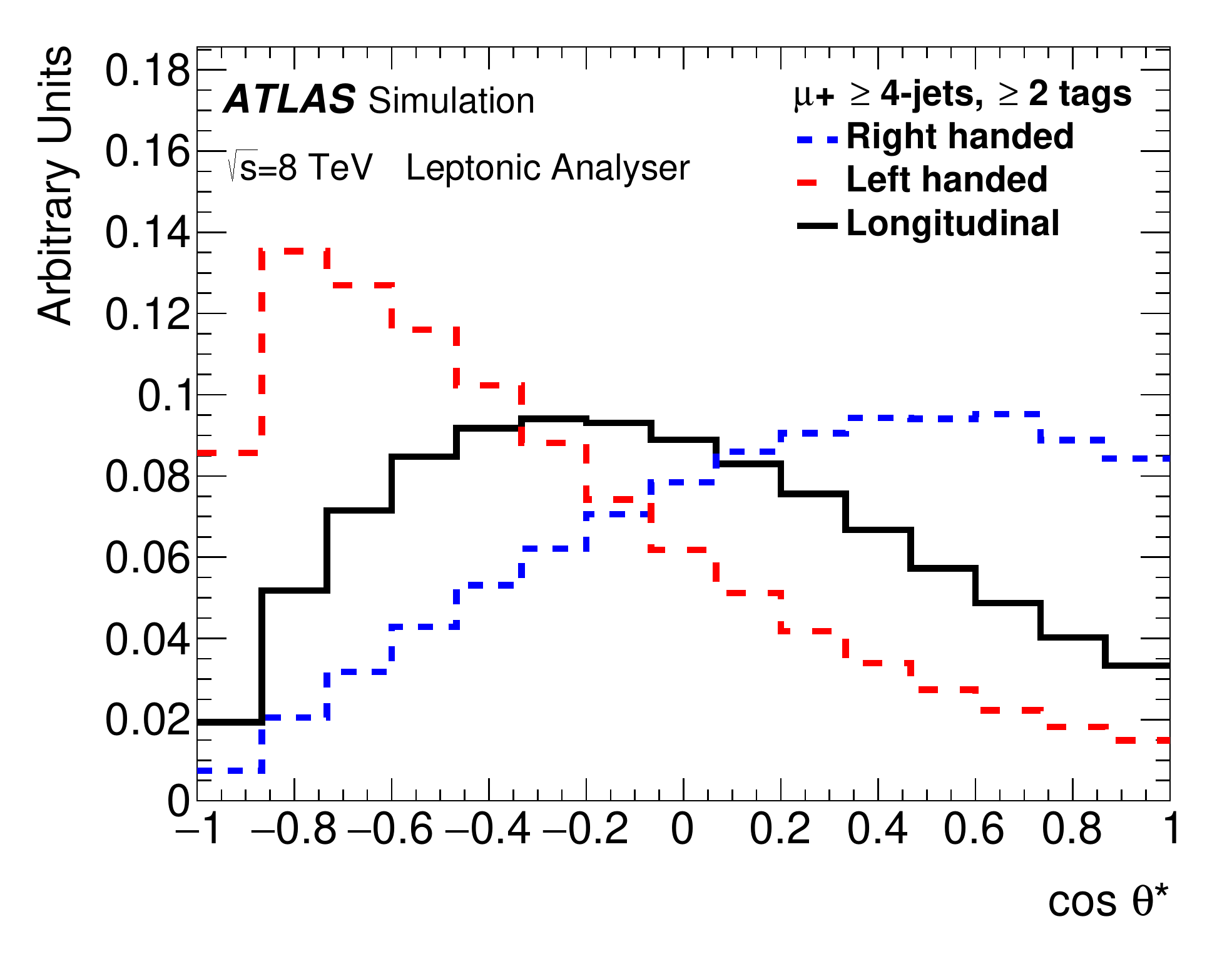}
}
\subfigure[]{
\includegraphics[width=0.42\linewidth]{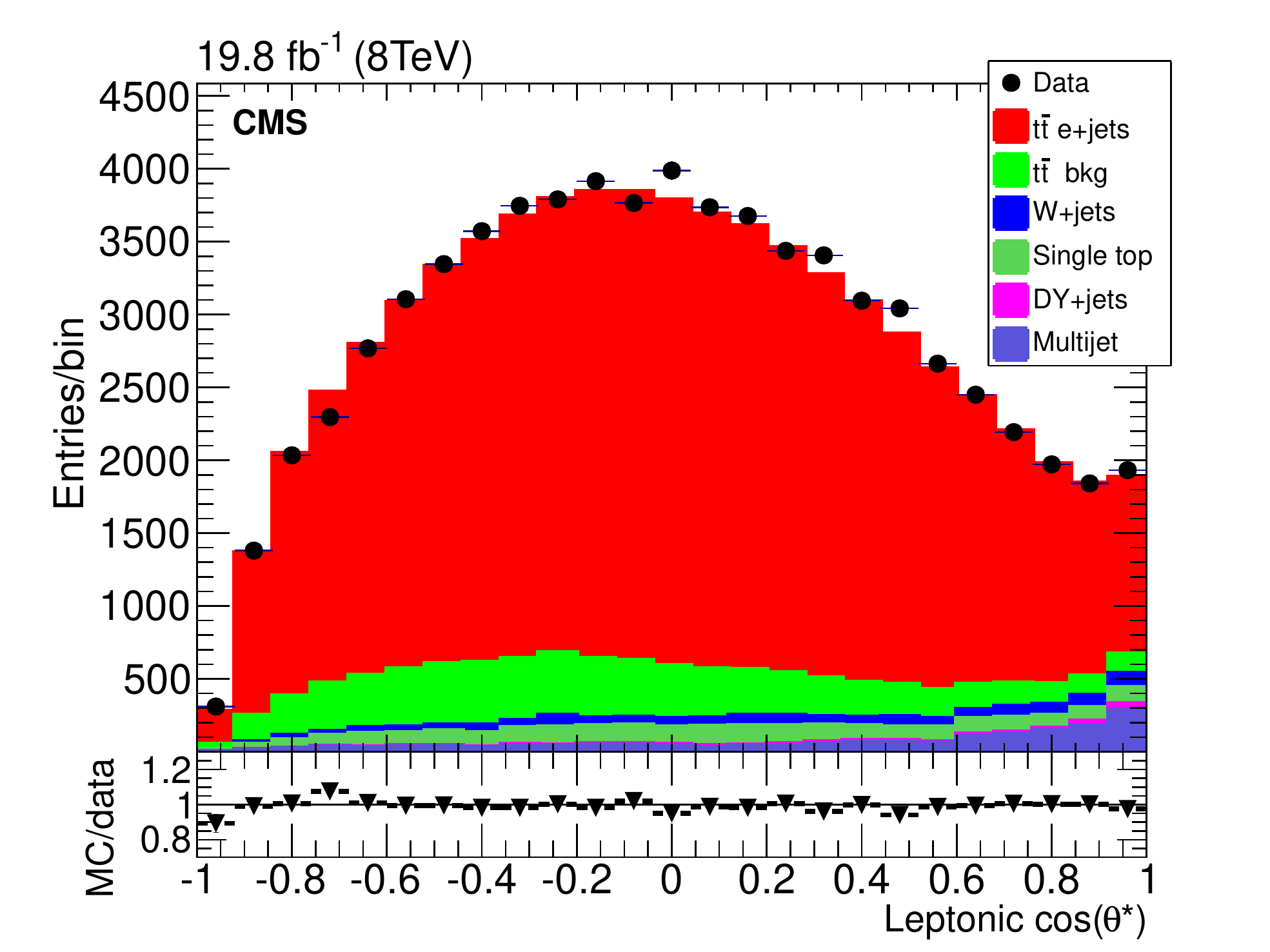}
}
\caption{Distribution of the cosine of the helicity angle ($\theta^*$) between the charged lepton and the direction of the top quark
in the $W$~boson rest frame. (a) Expected distributions for the left-handed, right-handed and longitudinal polarisation states in the ATLAS
$W$~polarisation measurement. (b) The CMS data are compared to the SM expectation.}
\label{f:whel:templates}
\end{figure}

\begin{figure}
\centering
\includegraphics[width=0.7\linewidth]{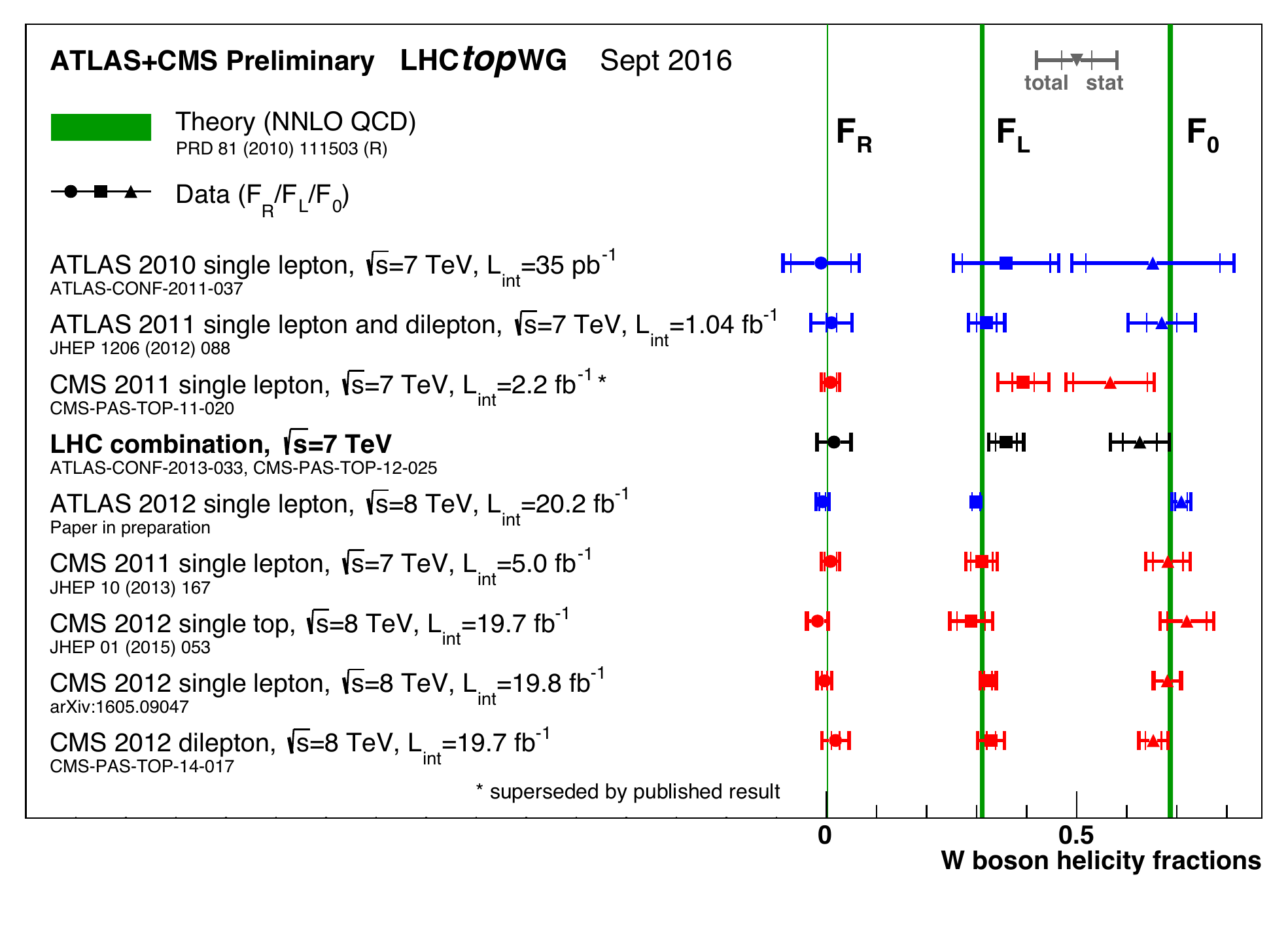}
\caption{Summary of LHC measurements of the $W$~boson polarisation fractions.}
\label{f:whel:summary}
\end{figure}

The ATLAS and CMS collaborations have both published recent measurements of the $W$~boson
polarisation fractions using the 8 TeV LHC data~\cite{Aaboud:2016hsq,Khachatryan:2016fky}.
Both experiments select events with one high transverse momentum ($\pt$) electron or muon
and at least four high $\pt$~jets. Kinematic fit techniques are then used to fully reconstruct
the $\ttbar$~system and the angle $\theta^*$~between the charged lepton and the direction
of the top quark is then reconstructed in the rest frame of the $W$~boson. The data are
fitted to the reconstructed $\cos \theta^*$~distributions in order to extract the measured
polarisation fractions. Figure~\ref{f:whel:templates} shows the templates in the ATLAS
analysis for the left-handed, right-handed and longitudinal polarisation states. There is clear discriminating
power between the different polarisation states.
The $\cos \theta^*$~distribution measured by CMS is shown in Figure~\ref{f:whel:templates} and the data are seen to be in good agreement with the SM predictions.
The largest uncertainties in the measurements arise from the Monte Carlo (MC) modelling of top quark events and
the jet energy scale (JES). The results are compared to previous LHC results and the SM prediction in Figure~\ref{f:whel:summary},
where good agreement between the SM and the measurements can be seen.

\subsection{Top quark polarisation}

Electroweak single-top quark production at the LHC is dominated by $t$-channel exchange
and the top quarks produced are predicted to be highly polarised, in particular along the direction
of the spectator-quark momentum~\cite{STpol1,STpol2}.
The polarisation ($P$) is related to the angle between a top-quark decay product and the top-quark spin axis ($\theta_{l}$) according to:
\begin{equation}
\frac{1}{\Gamma}\frac{d \Gamma}{d \cos \theta_{l}} = {\frac{1}{2}}(1 + \alpha P \cos \theta_{l}),
\end{equation}
where $\alpha$~is the spin analysing power of the decay product, which for charged leptons
is $\alpha(\ell^{\pm}) = \pm 0.998$~\cite{STpol3}.

ATLAS has recently measured~\cite{Aaboud:2017aqp} a set of angular asymmetries that are sensitive to the top-quark polarisation
and six independent $W$~boson spin observables~\cite{STpol4}. The measurements use events with one high $\pt$~electron or muon and exactly
two jets (one of which must be identified as originating from a $b$-quark). Selection requirements are imposed to reject the background
from $W$+jets and $\ttbar$~events.
The measured angular forward-backward asymmetries
\begin{equation}
A_{\mathrm{FB}} = \frac{N(\cos \theta > 0) - N(\cos \theta < 0)}{N(\cos\theta > 0) + N(\cos \theta < 0)}
\end{equation}
for two angles $\theta_{l}$ and $\theta_{l}^N$~are related to the top polarisation ($P$) and the $W$~boson spin 
observable $\langle S_2\rangle$~according to: $P = 2 \frac{A^{\ell}_{\mathrm{FB}}}{\alpha}$~and $\langle S_2\rangle = -\frac{4}{3} A^N_{\mathrm{FB}}$.
The values of the observables extracted from the asymmetries are $P = 0.97\pm 0.12$~and
$\langle S_2\rangle = 0.06 \pm 0.05$. Good agreement is seen between the data and the SM predictions.
The CMS experiment has previously measured the polarisation in single top events, finding $P=0.52\pm0.22$~\cite{Khachatryan:2015dzz}.
The measurement
agrees within two standard deviations with the SM prediction of $0.9$.

\begin{figure}
\centering
\subfigure[]{
\includegraphics[width=0.42\linewidth]{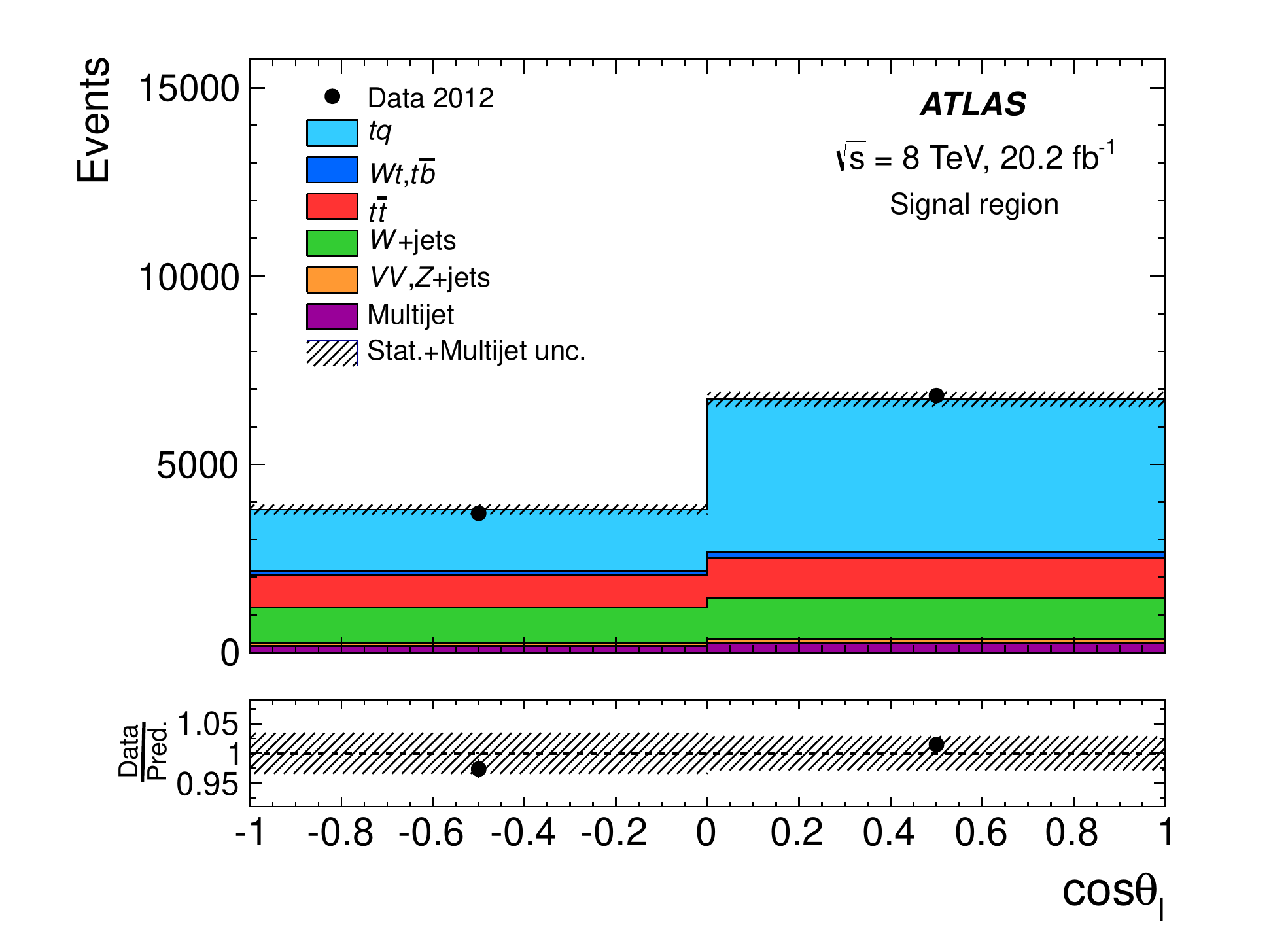}
}
\subfigure[]{
\includegraphics[width=0.42\linewidth]{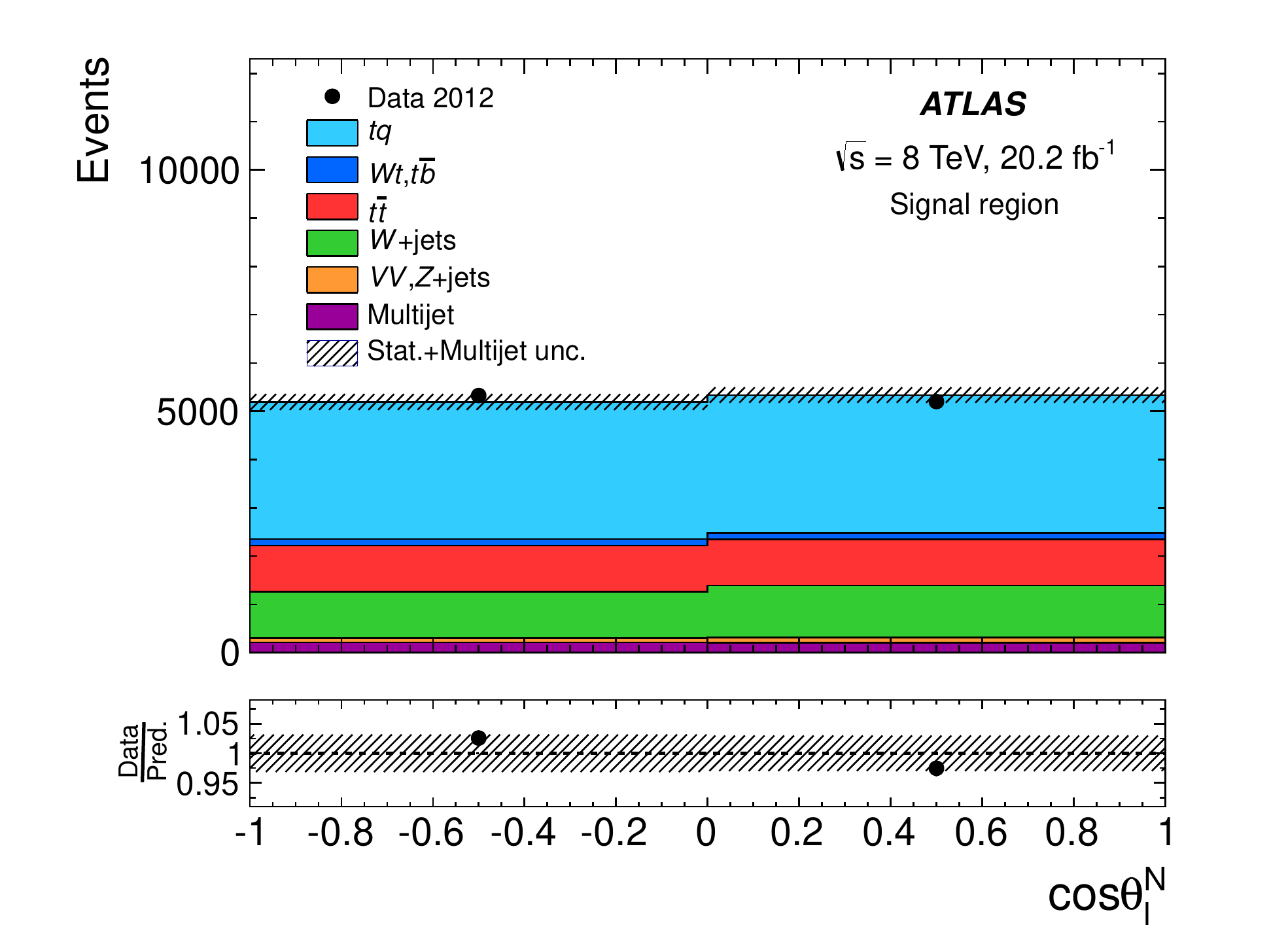}
}
\caption{Distributions of the cosine of the angles (a) $\theta_l$, which is sensitive to the top-quark polarisation and (b) $\theta^N_l$, which is sensitive to the spin
observable $\langle S_2\rangle$. The data (black points) are compared to the expectation of SM single-top production (light blue) and the background processes (other colours). }
\label{f:stpol:angles}
\end{figure}

\subsection{Constraints on the $Wtb$~vertex}

If the energy scale of new physics is not directly accessible in top quark production at the LHC, then the impacts of new physics
can be parameterised in the effective operator formalism and the most general $Wtb$~Langrangian can be written as:
\begin{equation}
{\cal L}_{Wtb} = - \frac{g}{\sqrt{2}}\,{\overline{b}}\gamma^\mu \left (\vl \ProjL + \vr \ProjR \right )tW^-_\mu 
- \frac{g}{\sqrt{2}}\,{\overline{b}}\,\frac{i\sigma^{\mu\nu}q_{\nu}}{\mW} \left (\gl \ProjL + \gr \ProjR \right)tW^-_\mu + \mathrm{h.c.}
\label{eq:lagrangian}
\end{equation}
The terms $\Vlr$ and $\Glr$ are the left- and right-handed 
vector and tensor couplings, respectively.\footnote{
In Equation~\ref{eq:lagrangian}, $g$~is the weak coupling constant, $\mW$ and
$q_{\nu}$ are the mass and the four-momentum of the $W$~boson,
respectively, $\ProjLR \equiv (1\mp \gamma^5)/2$ are the left- and
right-handed projection operators, and $\sigma^{\mu\nu} =[\gamma^{\mu}, \gamma^{\nu}]/2$.  
}
In the SM at tree-level, $V_{\mathrm{L}}$~is the CKM matrix element $V_{tb}$~and the anomalous couplings
$g_{\mathrm{L}}$, $V_{\mathrm{R}}$~and $g_{\mathrm{R}}$~are all zero.
The $W$~boson and top-quark polarisation measurements discussed in the previous sections
have been used to place limits on the anomalous couplings.

\begin{figure}
\centering
\subfigure[]{
\includegraphics[width=0.46\linewidth]{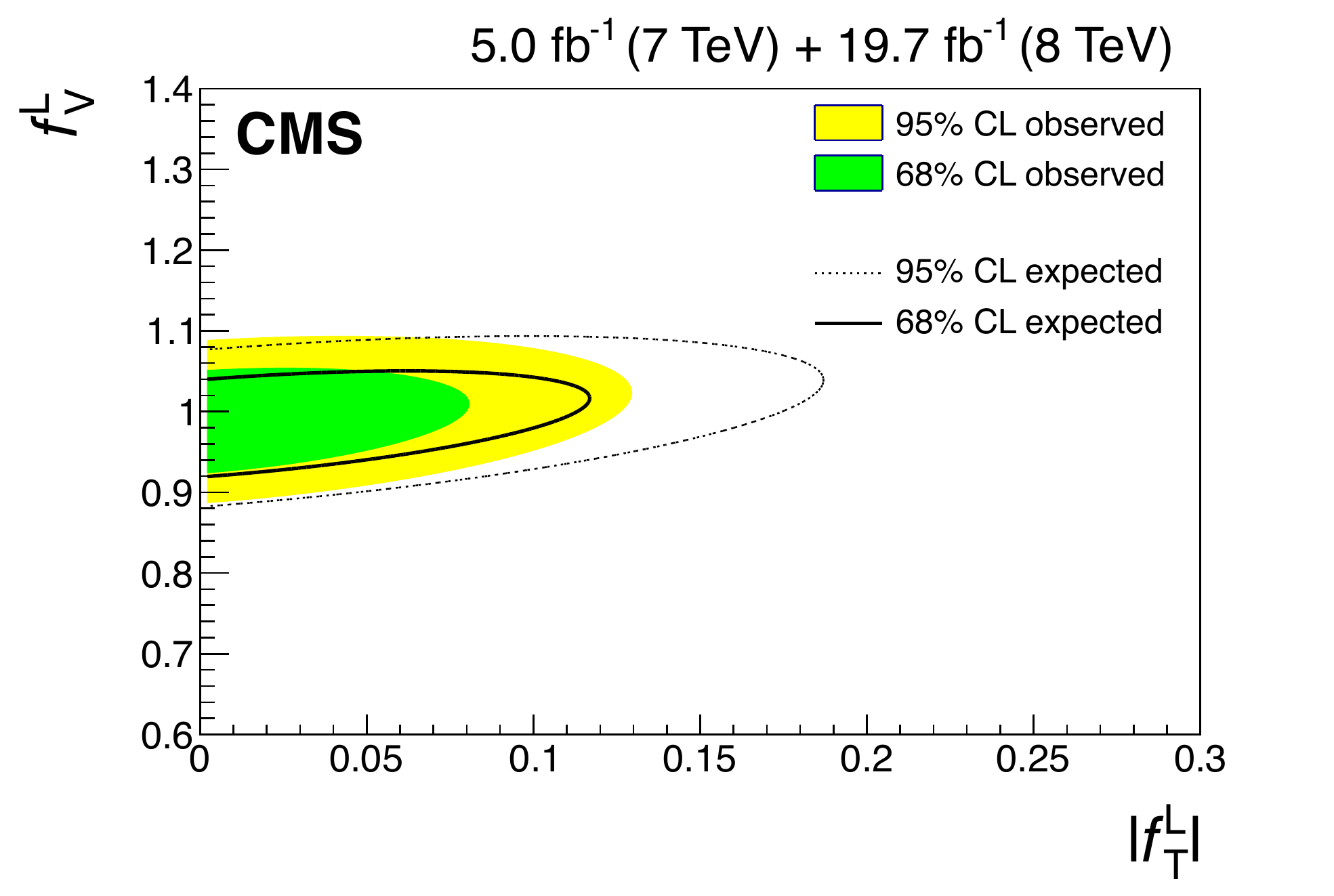}
}
\subfigure[]{
\includegraphics[width=0.42\linewidth]{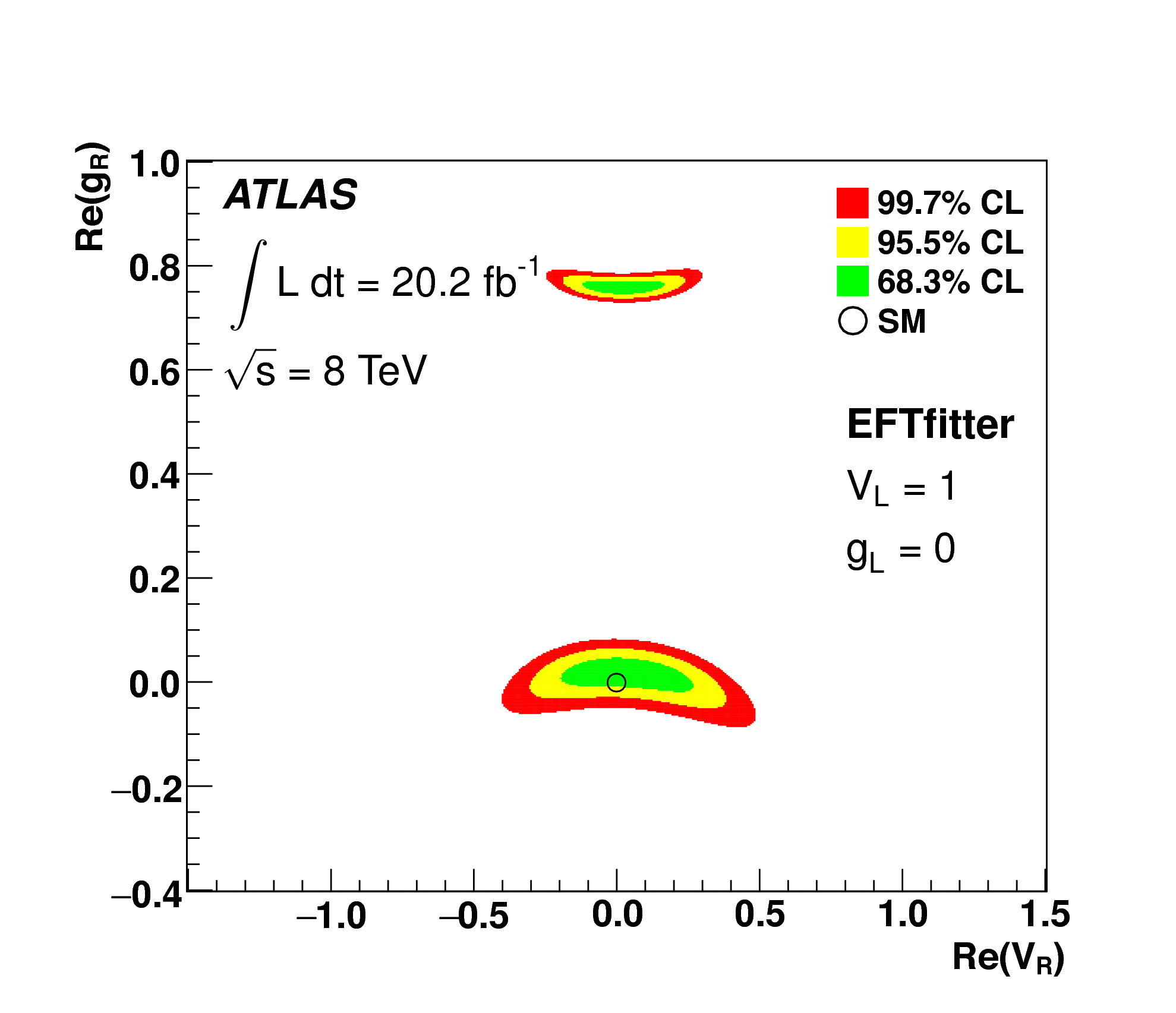}
}
\caption{(a) Limits on the anomalous couplings $\vl$~(labelled as $f^{\mathrm{L}}_{\mathrm{V}}$) and $\gl$~(labelled as $f^{\mathrm{L}}_{\mathrm{T}}$)
set by the CMS search using single-top events. (b) Limits on the anomalous couplings $\gr$~and $\vr$~set by the ATLAS measurement of the $W$-boson
polarisation fractions.}
\label{f:wtb:limits}
\end{figure}

The single top polarisation measurement is mainly sensitive to the imaginary part
of $g_{\mathrm{R}}$. The ATLAS measurements of the angular asymmetries for $\theta_l$~and $\theta^N_l$
are used in conjunction with analytical expressions~\cite{STpol4,STpol5,STpol6} to extract limits on $\mathrm{Im}\ g_{\mathrm{R}}$.
The correlation between the two asymmetry measurements $(-0.05)$~is accounted for in the limit setting procedure.
The limits set at the 95\% confidence level are Im\,\gr~$\in$~[$-$0.18, 0.06]. The CMS experiment
has designed a dedicated analysis to search for anomalous couplings in single-top events~\cite{Khachatryan:2016sib},
where multivariate classifiers are used to separate the SM single-top events from potential contributions
from non-zero anomalous couplings. No significant excess is seen and limits are set on different combinations of couplings.
Figure~\ref{f:wtb:limits} shows the limits set in the $\vl$~and $\gl$~plane.

The measurement of the $W$~boson polarisation fractions by ATLAS has been used to set limits
on the real parts of the anomalous couplings using the EFTfitter tool~\cite{EFTfitter}. 
Figure~\ref{f:wtb:limits}~shows the limits set on the \gr~and \vr~couplings, under the assumptions
$\vl=1$~and $\gl=0$.
The different precision measurements sensitive to the $Wtb$~vertex are complementary. In the interpretations
done by the collaborations to date, it has always been necessary to set at least one of the couplings to the SM
values. This motivates future combinations of these precision measurements, in order to obtain constraints
that are free of SM assumptions and to exploit the complementary sensitivity of the different measurements.

\section{Production of top quarks in association with vector bosons}

The large integrated luminosity delivered by the LHC allows the possibility to study the rare production of top-quark pairs in association
with either a $Z$~or $W$~boson ($\ttbar+V$). The production of $\ttbar+Z$~is particularly interesting, since it probes the top-$Z$~coupling.
ATLAS and CMS have both measured the $\ttbar+Z$~and $\ttbar+W$~cross-sections using the 13 TeV data~\cite{Aaboud:2016xve,CMS-PAS-TOP-16-017}.
The ATLAS measurement uses the 2015 dataset (corrsponding to a luminosity of $3.2$~fb$^{-1}$), while the CMS measurement uses the data collected
during the first half of 2016, which corresponds to an integrated luminosity of $12.9$~fb$^{-1}$. The measurements are limited by statistics and hence this
report will focus on the more precise CMS measurement.

The $\ttbar+V$~processes can produce final states with multiple-leptons and jets originating from $b$-quarks. Both experiments
select events with either two leptons (electrons or muons) with the same-sign charge, three leptons or four leptons.
For the dilepton channel, CMS selects events with at least 2 jets, at least one of which is identified as being likely to have originated from
a $b$-quark (referred to as a $b$-jet) and
then uses a multivariate technique to separate the signal from the $\ttbar$~background. To maximise the signal significance,
the dilepton events are further categorised according to the dilepton charge and the number of jets and $b$-jets.
The trilepton events are required to have at least 2 jets and events where a same-flavour
opposite-sign charge (SFOS) lepton pair has an invariant mass close to the Z boson mass are rejected.
The events are then categorised according to the number of jets and $b$-jets into twelve disjoint signal regions.
The tetralepton events are required to have one SFOS lepton pair consistent with a Z boson and to contain
at least two jets. In the $\mu\mu\mu\mu$, $eeee$~and $\mu\mu ee$~channels, events
where the second SFOS lepton pair is consistent with a Z boson are rejected to reduce the background from $ZZ$~events.
Events are categorised according to whether or not they contain at least one $b$-jet.

\begin{figure}[bhtp]
\centering
\subfigure[]{
\includegraphics[width=0.43\linewidth]{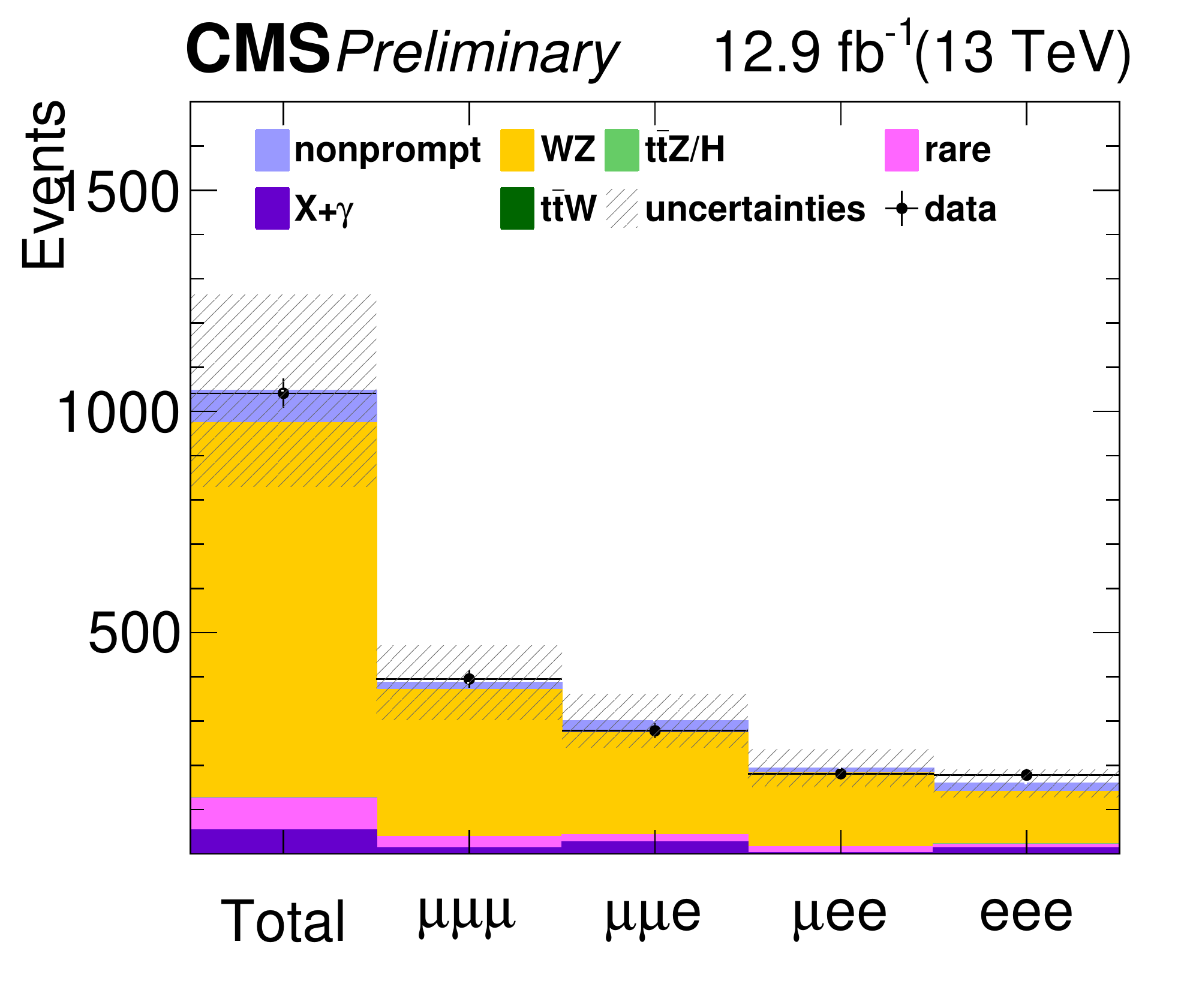}
}
\subfigure[]{
\includegraphics[width=0.38\linewidth]{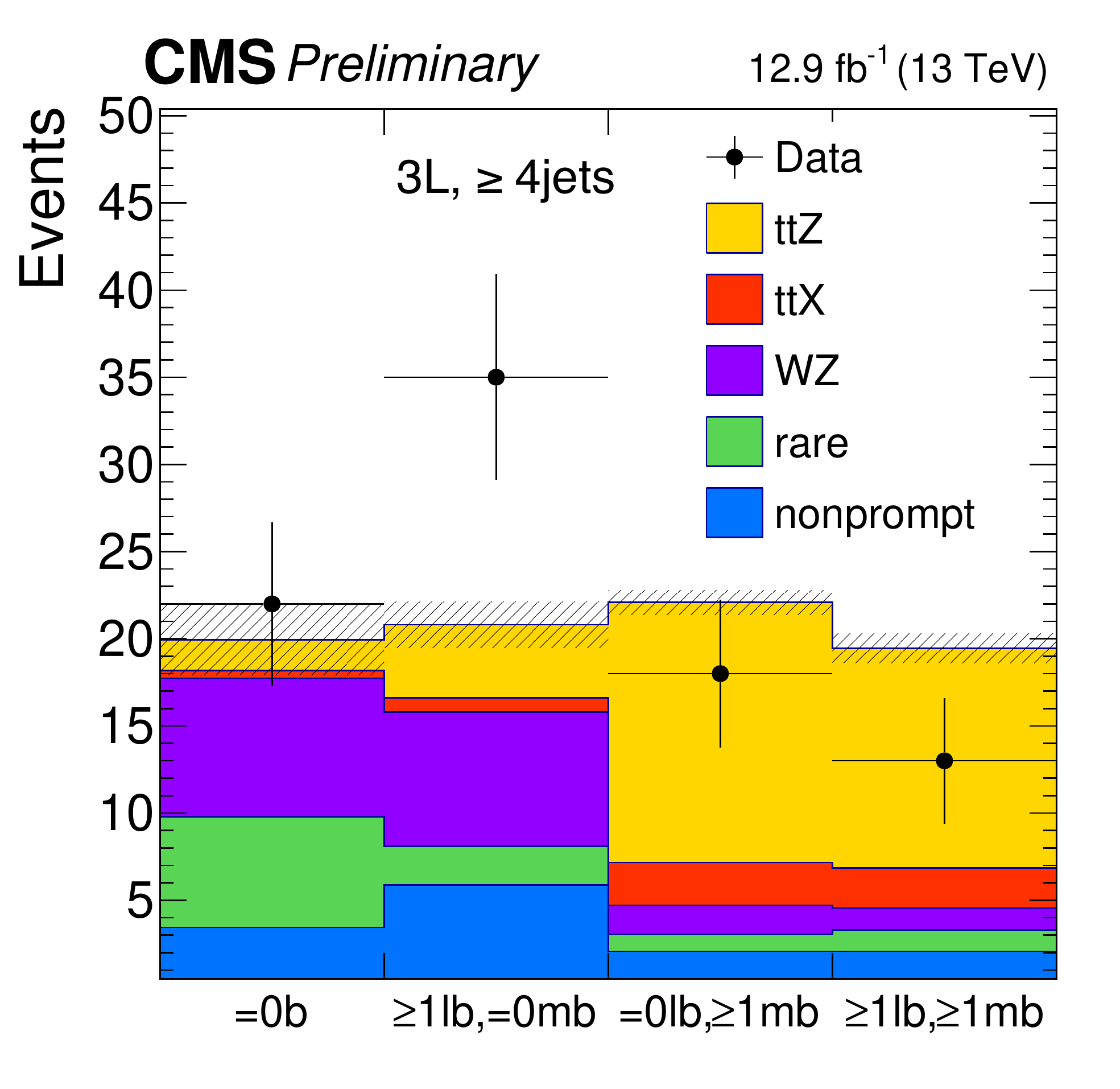}
}
\caption{(a) Number of events in each lepton flavour channel for the $WZ$~control region in the CMS $\ttbar+V$~measurement.
(b) Number of events in the different $b$-jet categories in the CMS trilepton $\ttbar+V$~signal region with at least four jets, where
$mb$~refers to the number of jets passing a $b$-tagging requirement with an efficiency of $70\%$~and $lb$~refers to the number of
jets failing the medium requirement and passing a loose requirement (with an efficiency of $85\%$).
In both cases the data points in black are compared to the sum of the expected SM processes.
}
\label{f:ttV:plots}
\end{figure}

The background predictions use
data-based methods for the the backgrounds where at least one lepton originates from non-prompt sources (leptons from heavy-flavour hadron decay,
misidentified hadrons, muons from light-meson decay in flight, or electrons from unidentified photon conversions), while other background sources
are estimated using simulation. The modelling of the backgrounds is checked in control regions. Figure~\ref{f:ttV:plots} shows
the number of events in each lepton flavour channel for a control region that selects events with three leptons and at most one jet.
This region is dominated by $WZ$~events and good agreement is seen between the data and the background prediction.

\begin{figure}
\centering
\subfigure[]{
\includegraphics[width=0.46\linewidth]{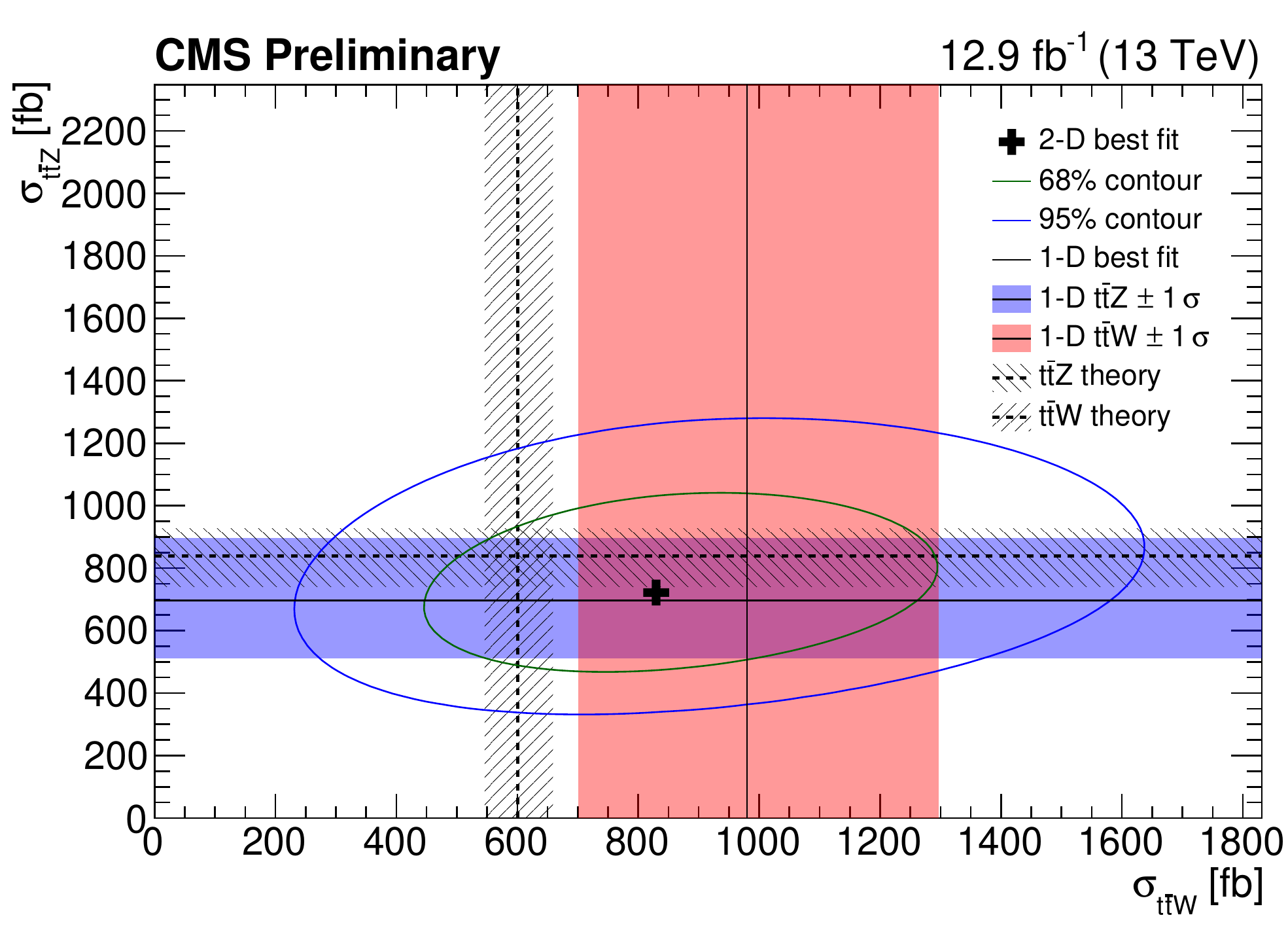}
}
\subfigure[]{
\includegraphics[width=0.42\linewidth]{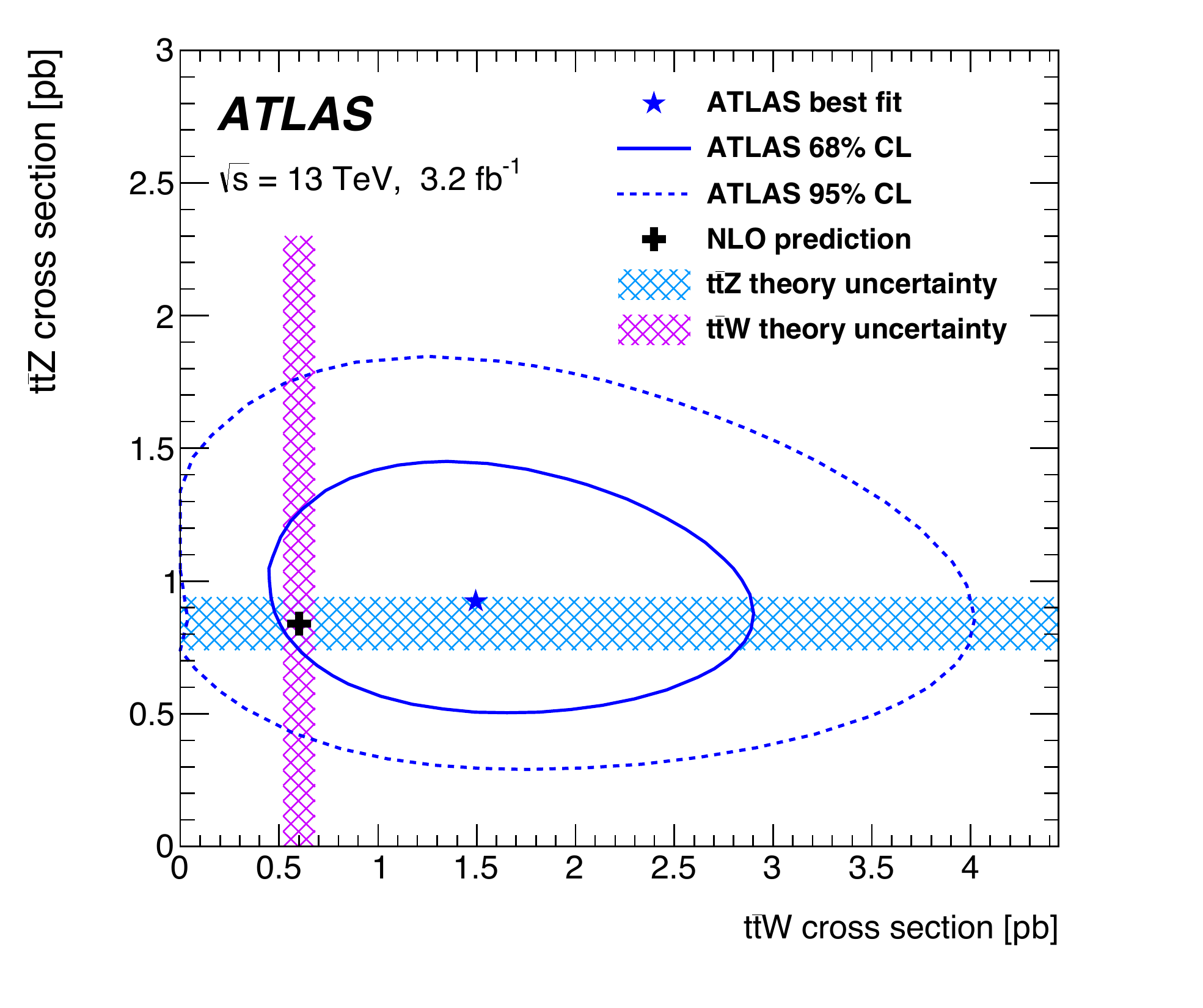}
}
\caption{The measured cross-sections for the $\ttbar+W$~and $\ttbar+Z$~cross-sections are compared to the theoretical
predictions for (a) the CMS results and (b) the ATLAS results. 
}
\label{f:ttV:xs}
\end{figure}

The $\ttbar+W$~and $\ttbar+Z$~cross-sections are extracted by making a combined fit to all the different signal regions.
The data observed in the trilepton signal region with at least four jets are shown in Figure~\ref{f:ttV:plots}, where
good agreement is seen between the data and the SM prediction. The measured cross-sections
are shown in Figure~\ref{f:ttV:xs} for both the ATLAS and CMS analyses. Agreement is seen between the data
and the SM predictions. The measurements are dominated by the statistical uncertainty and hence measurements
utilising larger datasets are eagerly awaited.

\section{The top quark mass}

The top quark mass is a fundamental parameter in the SM and precisely measuring its value
is vital for establishing the consistency of the SM~\cite{Baak:2014ora}. ATLAS has recently performed
a measurement using top-quark pair events in the dilepton channel~\cite{Aaboud:2016igd}.
The events are selected from the 2012 data taken at $\sqrt{s}=8$~TeV and the dataset corresponds
to an integrated luminosity of $20.2$~fb$^{-1}$. The selection requires exactly two
leptons (either electrons or muons) and at least two jets, where at least one of the jets is required
to be a $b$-jet. Additional requirements on the missing transverse momentum, the invariant mass of the lepton
pair and the scalar sum of the $\pt$~of the selected jets and leptons are applied to reduce the background
from $Z$+jets events.
The dilepton channel has the drawback that the prescence of two neutrinos makes
the full reconstruction of the $\ttbar$~system challenging.
The two jets with the highest probability to originate from $b$-quarks are taken as originating
from the two top quarks. There are then two possible assignments of the leptons and $b$-jets.
The combination that leads to the lowest average invariant mass of the two lepton-$b$-jet pairs ($\mlb$)
is selected.
The top mass can then be extracted from the $\mlb$~distribution.
A phase-space restriction on the average $\pt$~of the two lepton-$b$-jet pairs ($p_{\mathrm{T},\ell b}$)
is used to obtain the smallest total uncertainty in $\mt$. The selected requirement is $p_{\mathrm{T},\ell b} > 120$~GeV
and effectively selects a region where the systematic uncertainties are reduced.

\begin{figure}[htbp]
\centering
\subfigure[]{
\includegraphics[width=0.38\linewidth]{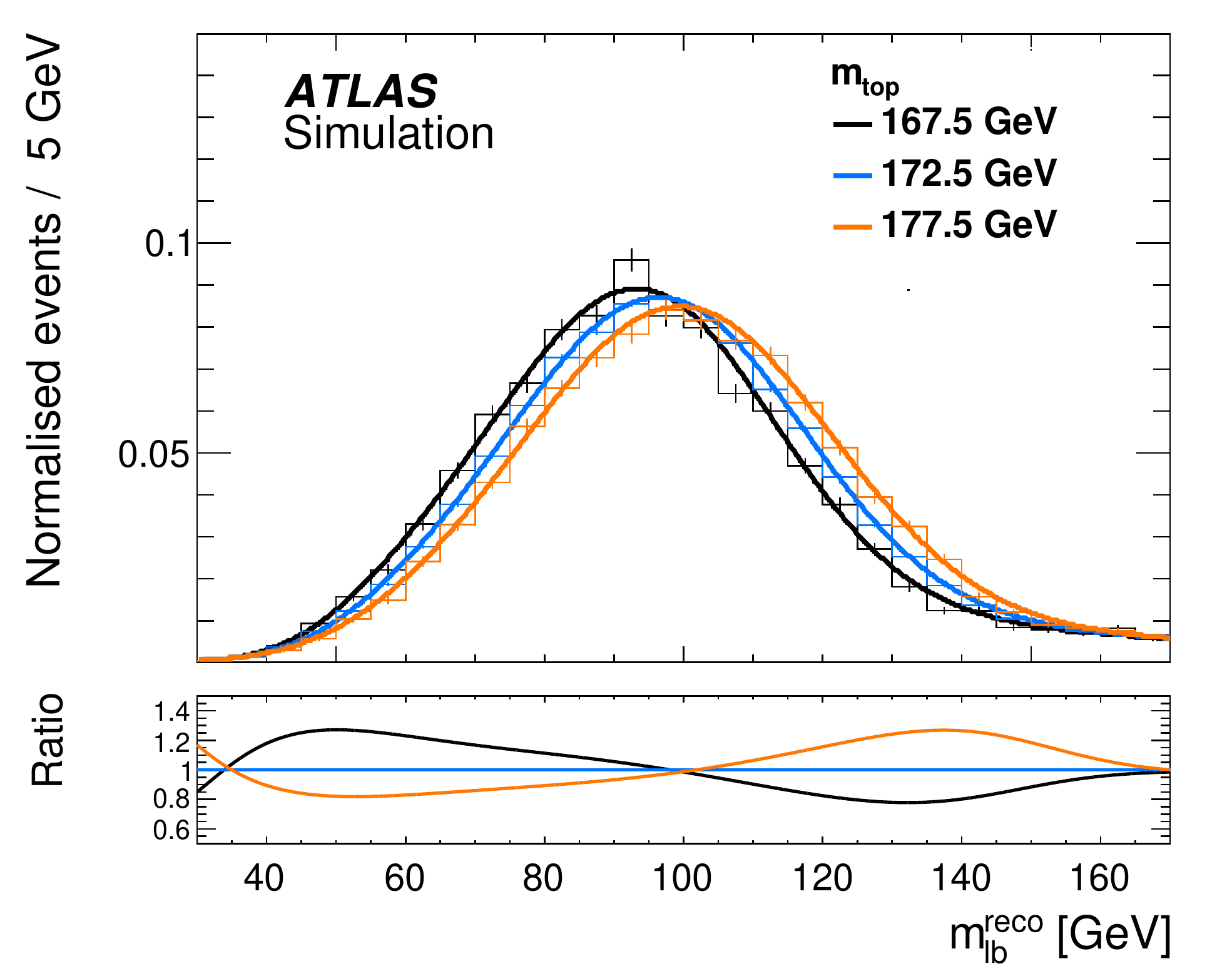}
}
\subfigure[]{
\includegraphics[width=0.38\linewidth]{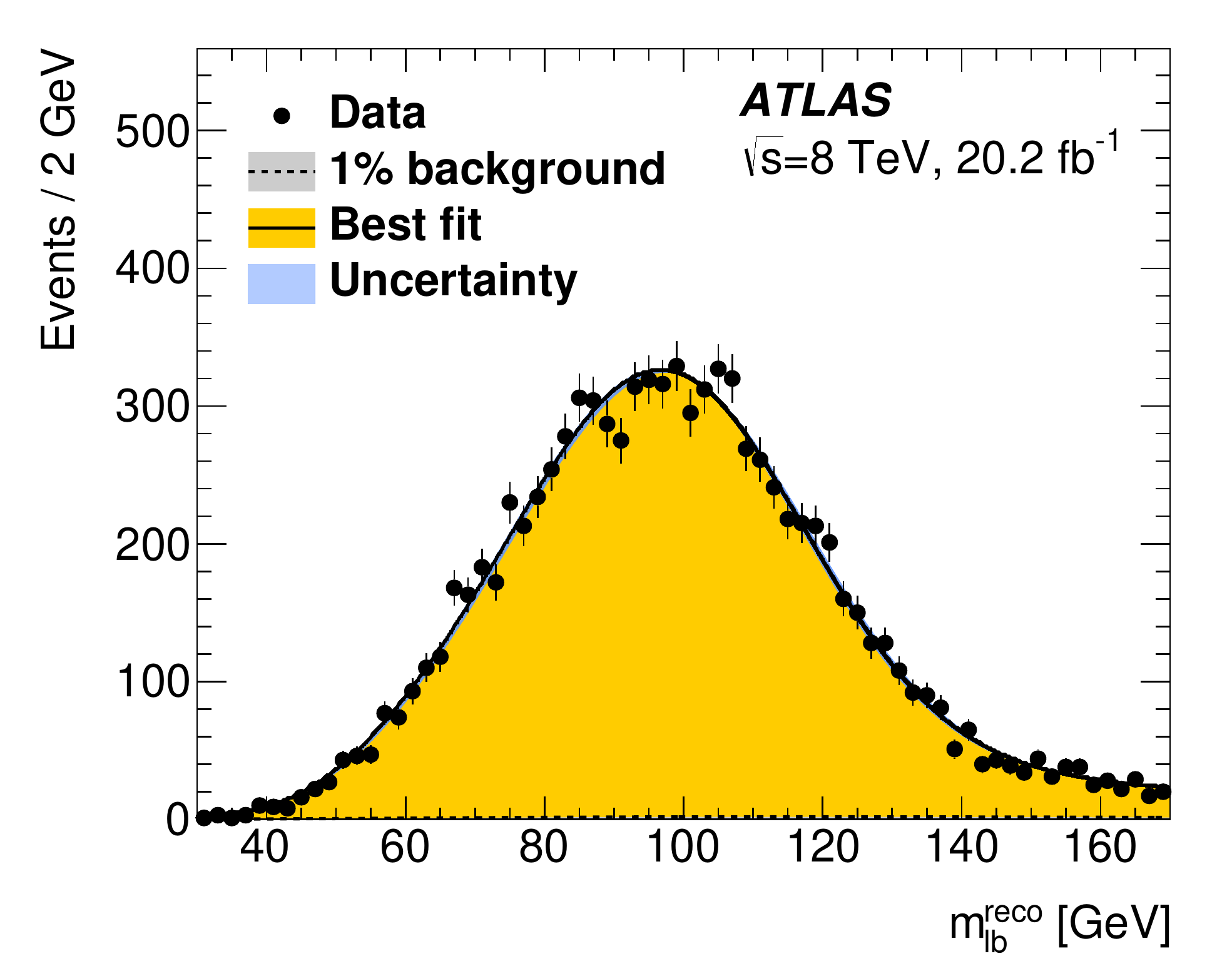}
}
\caption{(a) Distribution of the $\mlb$~variable for templates and MC samples with different \mt~values. (b) The template with the best fit for \mt~is compared
with the data (black points).
}
\label{f:atlasmtop}
\end{figure}

The top-quark mass is extracted by performing a template fit to the $\mlb$~distribution.
The signal templates are constructed by fitting Monte Carlo samples generated with different top quark masses
with the sum of a Landau function and a Gaussian distribution.
Figure~\ref{f:atlasmtop} shows the templates for three different $\mt$~values, demonstrating the sensitivity of
the $\mlb$~distribution to the top-quark mass. The figure also shows the data compared to the template
with the best fit value of the top-quark mass and good agreement is seen between the data and the fitted template.
The top-quark mass is measured to be $\mt = 172.99 \pm 0.41 \pm 0.74$~GeV, where the first uncertainty is
the statistical uncertainty and the second is the total systematic uncertainty. The systematic uncertainty
is dominated by the understanding of the jet energy scale ($0.54$~GeV), the MC modelling of top-quark pair events ($0.35$~GeV)
and the jet energy scale for jets originating from $b$-quarks ($0.30$~GeV). The measurement is the most precise measurement
of the top-quark mass in dilepton events to date.

\begin{figure}[htbp]
\centering
\subfigure[]{
\includegraphics[width=0.38\linewidth]{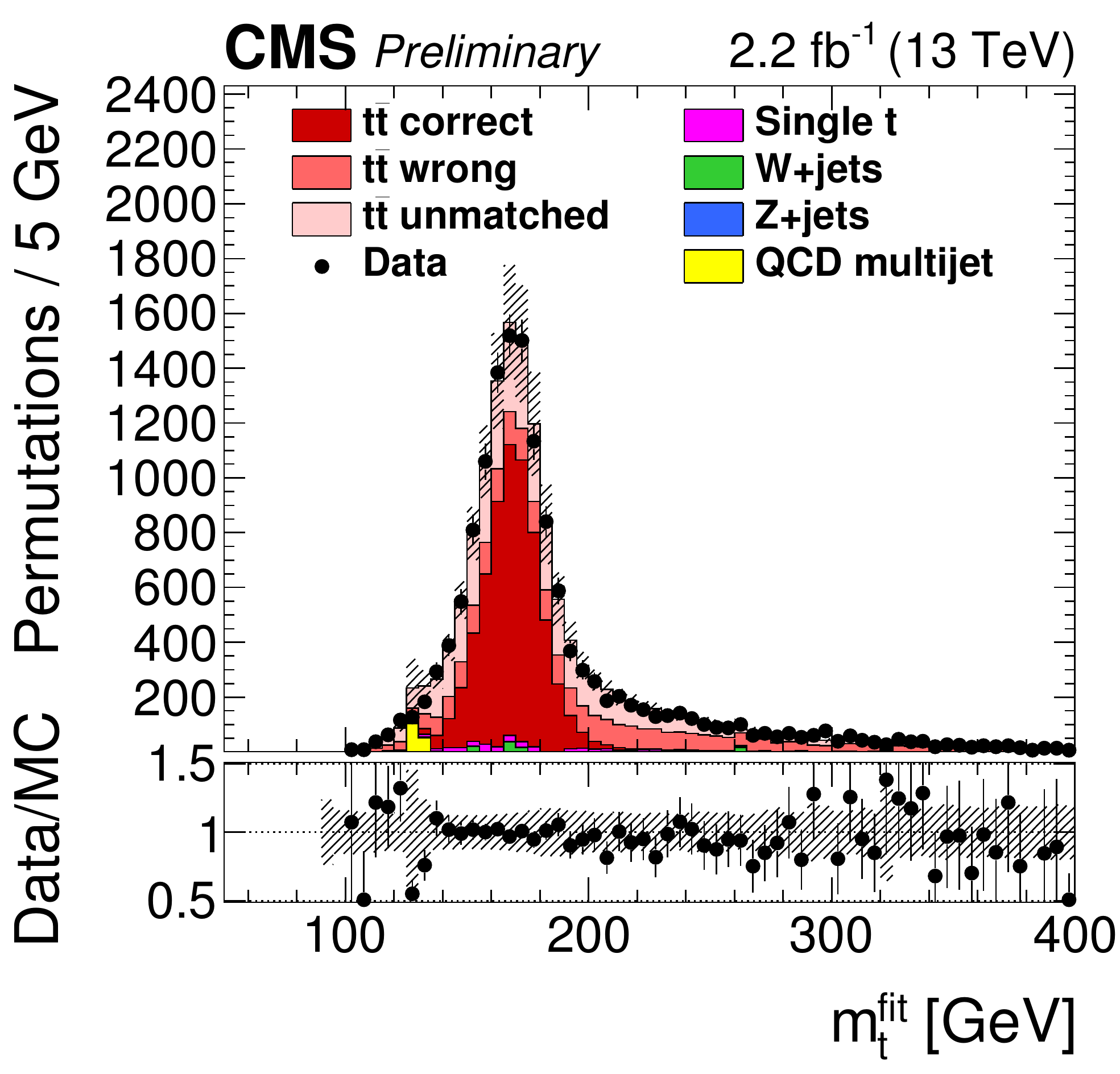}
}
\subfigure[]{
\includegraphics[width=0.38\linewidth]{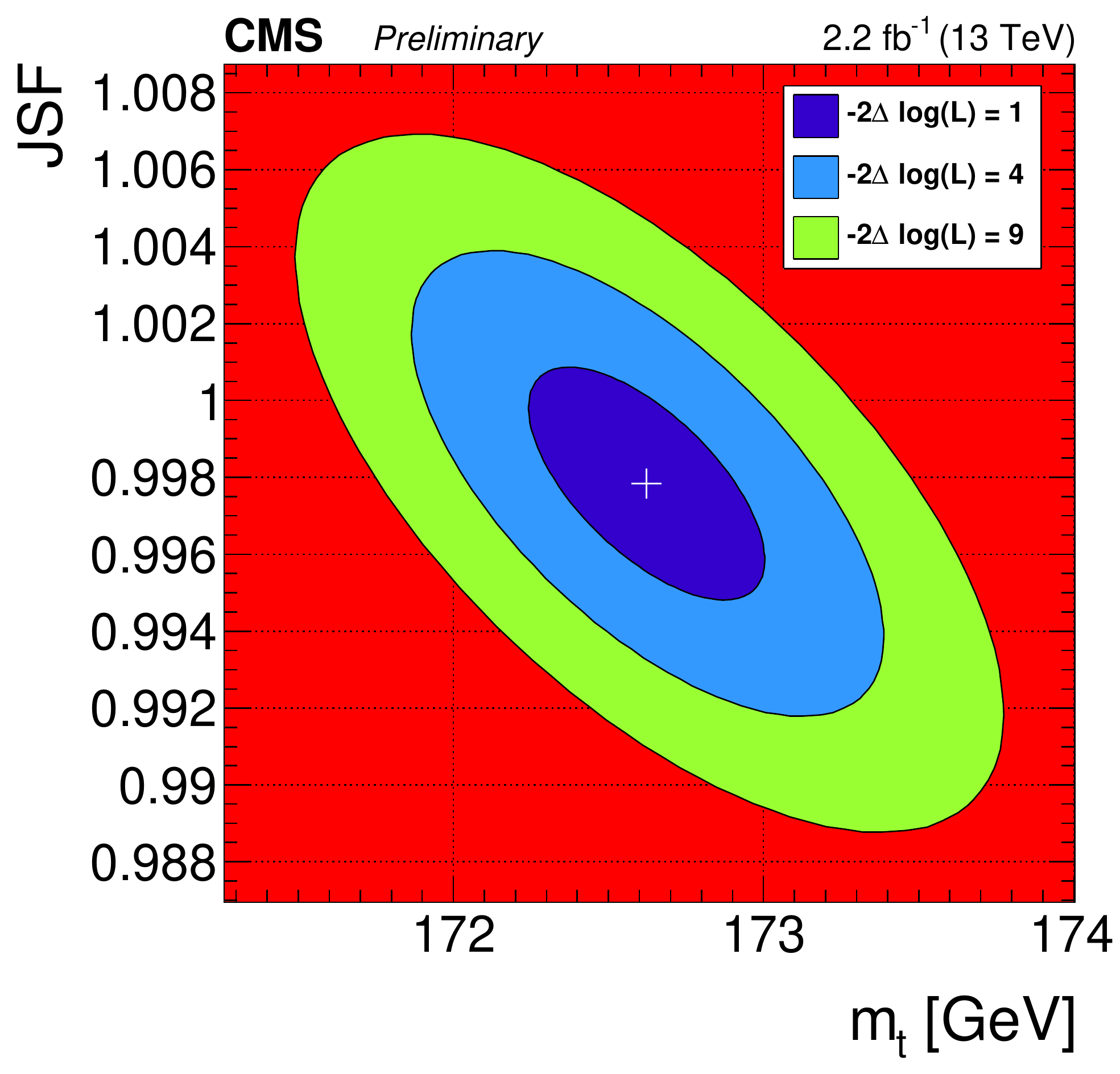}
}
\caption{(a) Distribution of the reconstructed top-quark mass is compared to the simulation. (b) Fitted values for the top-quark
mass and the JSF; the coloured contours show the statistical uncertainty.
}
\label{f:cmsmtop}
\end{figure}

The CMS collaboration has recently made the first measurement of the top-quark mass using the 13 TeV LHC data~\cite{CMS-PAS-TOP-16-022}.
The analysis uses events with one muon and at least four jets (of which two must be $b$-jets)
and the strategy closely follows the lepton+jets analysis that used the run 1 data~\cite{Khachatryan:2015hba}.
A kinematic fit is applied to the selected events, where the fit constrains the
$W$~boson mass to $80.4$~GeV~\cite{pdg} and the top and anti-top masses to be the same.
The goodness-of-fit probability (gof) of the kinematic fit is required to be at least 0.2 in order to remove
poorly reconstructed events. 
The top mass reconstructed from the kinematic fit ($m_{\mathrm top}^{\mathrm fit}$) is used along
with the reconstructed $W$~boson mass in a likelihood-fit to simultaneously extract
the top-quark mass and the jet energy scale factor (JSF). The JSF allows the overall energy scale
of jets to be constrained using the reconstructed $W$~boson mass.
In the likelihood-fit, events are weighted by their gof in order to reduce the impact of poorly reconstructed events.
The reconstructed top mass distribution is shown in Figure~\ref{f:cmsmtop}, which shows the excellent
mass resolution and good agreement between the data and the simulation.
The top quark mass is measured to be $\mt = 172.62 \pm 0.38 \pm 0.7$~GeV, where
the first uncertainty is statistical and the second is the total systematic uncertainty.
The fitted JSF is in agreement with 1, as shown in Figure~\ref{f:cmsmtop}.
The largest systematic uncertainties are from the jet energy scale ($0.51$~GeV) and
the MC modelling of top-quark events ($0.4$~GeV). While the total uncertainty does not reach
the precision achieved in run 1~\cite{Khachatryan:2015hba}, the understanding of the 13 TeV data
is still at an early stage compared to run 1 and improvements in the uncertainties in the future
can be anticipated.

\section{Summary}

The heavy mass of the top quark provides the opportunity to make precision measurements of a
`bare' quark. Recent measurements of $W$~boson polarisation fractions in top-quark pair events
and of top-quark polarisation in single-top quark events have probed the structure of the $Wtb$~vertex.
The measurements are all in agreement with the SM, with no signs of new physics.
The high energy and large datasets provided by the LHC allow to measure the rare
$\ttbar+Z$~and $\ttbar+W$~processes. The recent CMS measurement with 13 TeV data
shows evidence for both processes and the larger datasets expected in the years ahead will
allow for precision tests of the top-$Z$\ coupling. The top-quark mass has been measured
to high precision with the LHC data. The recent dilepton measurement from ATLAS
utilises a kinematic phase space selection to reduce the systematic uncertainties and CMS
has made the first run 2 top mass measurement. Future prospects include a combination of the
LHC run 1 results and precise measurements with the run 2 data.

\end{document}